\documentclass[useAMS,usenatbib,amslatex]{mn2e}
\usepackage{graphicx,graphics,amsmath,amssymb,mathrsfs,textcomp}
\usepackage[latin9]{inputenc}
\usepackage{setspace}  
\usepackage{fancybox}
\usepackage[usenames,dvipsnames]{xcolor}
\usepackage{tikz}
\usetikzlibrary{shapes,shadows,arrows}

\newcommand{\dmath}{\mbox{${\mathrm d}$}}

\newcommand{\intgbth}[2]{\int_{#1}^{#2}}

\xdefinecolor{darkred}{rgb}{1.0,0.0,0.0} 

\def\mnras{{Mon.\@ Not.\@ Roy.\@ Ast.\@ Soc.}}
\def\aap{{Astron.\@ Astrophys.}}

\def\physrep{{Phys.\@ Rep.}}

\usepackage{subfigure}

\begin{document}
\title[Ram pressure stripping]{Ram pressure stripping: An analytical approach}
\author[Singh, Gulati and Bagla]{Ankit Singh$^1$\thanks{E-Mail: ankitsingh@iisermohali.ac.in},
  Mamta Gulati$^2$\thanks{E-Mail: mamta.gulati@thapar.edu} and 
Jasjeet S. Bagla$^1$\thanks{E-Mail: jasjeet@iisermohali.ac.in}\\
$^1$Indian Institute of Science Education and Research Mohali,
Knowledge City, Sector 81, Sahibzada Ajit Singh Nagar, Punjab 140306,
India \\
$^2$ School of Mathematics, Thapar Institute of Engineering and Technology, Patiala, Punjab 147004, India\\}
\maketitle

\begin{abstract}
  We use an analytical approach to study ram pressure stripping with
  simple models for discs and halo gas distribution to study the
  phenomena in cluster, group and galaxy halos. 
  We also study variations with galaxy properties and redshift.
  In each case we model the worst case scenario (i.e., maximum effect due
  to ram pressure). We show that there is little variation in the worst 
  case scenario with redshift.
  We find that gas discs in galaxies with a higher spin parameter get
  stripped sooner than galaxies with a smaller spin parameter.
  Galaxies in cluster halos get stripped of gas more efficiently as
  compared to group and galaxy halos: this is due to the higher infall
  speed and a higher density of gas in the ICM due to a greater
  retention of baryons.  
  We comment on the limitations of our model and situations where
  a significant amount of gas may be retained in galaxy disc and 
  also give an illustration for the same.
  Lastly, we discuss implications for star formation in galaxies as
  these fall into halos. 
\end{abstract}


\section{Introduction}

Observations point to a lack of gas-rich galaxies in clusters. 
To explain the absence of gas-rich galaxies in the clusters,
\citet{GG72} proposed the mechanism of ram pressure stripping.
A galaxy falling in a cluster experiences a wind as it moves through
the Intra-Cluster Medium (ICM). 
If the gravitational force between the stellar and gas disc is less
than the force due to the wind, then the gas will be stripped.
The condition for gas loss is given by:
\begin{equation}
\rho^{}_{\scriptscriptstyle ICM}V_{\rm in}^{2} > 2\pi G \Sigma_{s}\Sigma_{g}.
\end{equation}
Here, the right-hand side is a gravitational restoring pressure with
$\Sigma_{s}$ being the surface density of stars and $\Sigma_{g}$ is
the surface density of the gas, the density of ICM  is
$\rho^{}_{\scriptscriptstyle ICM}$, and the in-fall velocity of the
galaxy is given by $V_{\rm in}$.  
This condition, on putting numbers for spirals galaxies, led them to
conclude that a spiral galaxy should lose its gas disc while passing
through the centre of a cluster. 

It is observed
\citep{2003ApJG, 2003ApJH, 2006ApJG, 2008A&AV, 2009ApJP, 2010MNRASV, 2012MNRASW, 2016MNRASF}
that cluster centres contain galaxies with early-type morphologies
and have star formation rates that are much lower than that of field
galaxies with a similar luminosity.
Low Star Formation Rates (SFR) is due to deficiency of cold gas in
these galaxies \citep{2006A&AG}.
Observations of different clusters \citep{1983AJG, 2000AJB, 2001ApJS, 
jnc2001}
have shown the presence of lenticulars and HI deficient spirals.
\citet{2001ApJS, lah2009} observe an increase in deficiency with decreasing
distance from cluster centre and that the galaxies which are HI
deficient tend to have highly eccentric orbits.

\citet{1994AJC} observe signatures of gas stripping in galaxies in
Virgo cluster.
The inner discs of the galaxies have normal gas densities but are
sharply truncated when compared to field galaxies of similar
morphology.
In a study of more Virgo spirals, \citet{1999AJK,2004AJK} observe 
NGC$4522$ to have heavily truncated HI disc with one side containing
extraplanar H$\alpha$ and HI, but the stellar disc is undisturbed.
\citet{2004IAUSK} observe IC$3392$, NGC$4402$, NGC$4388$ and NGC$4419$
to have a truncated HI disc.
An undisturbed optical disc with an extraplanar gas component in
these galaxies indicates interaction of galaxy with the ICM.  
Interestingly one finds a positive correlation between the star
formation of the central galaxy and its satellite for the regions
extending far beyond virial radius \citep{2006MNRASW,2008MNRASA,2011Guo}, a
phenomenon known as 'galactic conformity'.
\citet{2017MNRASB} performed a multi-wavelength study taking data from
Sloan Digital Sky-Survey (SDSS) and Arecibo Legacy Fast ALFA (ALFALFA)
survey.  They observe that gas suppression in a high-density environment 
begins at the group level.
They conclude that the amount of gas suppression cannot be because of
a slow process like starvation or strangulation \citep{1980ApJL} and a
fast process like ram pressure stripping is needed to explain the
observation. Recently \citet{treyer2018} have shown low star formation 
activity in the group environment at low redshift owing to the environmental 
quenching processes. It is believed that the stripping acts at two levels 
in the group enviornment.
First, the dwarf galaxies with lower mass having lower restoring force
may be stripped. 
Second, the massive spirals may lose outer parts of their gas discs
as the restoring force is low in these regions.

\citet{boselli2018} observed star-forming regions up to 20 kpc 
outside the optical disc of the galaxy NGC$4285$. These regions are located 
along with a tail of HI gas stripped from the disc of the galaxy. Such 
galaxies are called jellyfish galaxies. \citet{jaff2018}, in 
a recent compilation of jellyfish galaxies in low-redshift clusters from
the WINGS and OmegaWINGS surveys, and follow-up MUSE observations from the
GASP MUSE programme investigated the orbital histories of jellyfish galaxies in
clusters and reconstructed their stripping history. They found that the 
jellyfish  galaxies seen in clusters are likely formed via fast, incremental, 
outside-in ram-pressure stripping during first infall into the cluster in
highly radial orbits. On-going ram pressure stripping in three new
intermediate-mass irregular galaxies in the north-west outskirts of
the nearby cluster of galaxies Abell 1656 (Coma cluster) is shown by
\citet{gavazzi2018}.

Systems with a galaxy passing through the medium have been simulated
to confirm the Gunn and Gott condition for gas stripping.
Simulations studying ram pressure have been done in broadly three
types of settings: wind tunnel   
\citep{1976A&AG,1994ApJB,2000SciQ,2009ApJT,2013MNRASC,2014MNRASW,2015A&AB},
simple systems  
\citep{2007MNRASR,2008MNRASK,2009A&AK} and cosmological simulations 
\citep{2001ApJV,2007ApJT,simpson2018}.
\citet{2005A&AR} in their simulation with lower ram pressure observe
that large and medium-sized model spirals have their outer HI disc
stripped.
\citet{2003MNRASM} in their simulation of a dwarf-like galaxy in the
group environment found that it will be stripped of its gas.
The observational evidence of such a system is circumstantial.
In case of local group dwarfs \citet{2000ApJB} observed that the star
formation history correlates well with their distances from the
dominant spirals. 

Morphological transformation of galaxies due to ram pressure is seen
in simulations, though studies indicate that this in itself is not
adequate to explain the observations \citep{1999MNRASA}.
The process of stripping is also found to have two stages. 
The first stage is rapid stripping of outer parts of the gas disc due
to the ram pressure and in the second stage the rest of the gas is
slowly stripped viscously \citep{2000SciQ,2001MNRASS,2003MNRASM,2005A&AR}. 
\citet{2005A&AR} performed simulations to study the dependence of the
stripping radius on the speed of ICM wind velocity and vertical
structure of the galaxy.
Their results confirm that the most important factors are the surface
density of the disc and the ram pressure.
The vertical structure of the disc and Mach number have a little role
to play in stripping.
The Gunn and Gott condition predicts the stripping radius fairly
well. 

While both simulations and observations seem to strongly suggest that
ram pressure stripping is a key quenching mechanism in the cluster
environment, along with few relevant studies for groups, its role in the 
group and galactic environments is still not clear. 
Although there is strong evidence for the presence of ram pressure
stripping, the observations also
seem to suggest that the observed relation of morphology and
the environment may be contaminated by the presence of other processes as
well.
Since the galaxies residing in the high-density environments formed
earlier than their field counterpart having same luminosity, they are
expected to be redder and have a lower star formation.
Moreover, galaxies in high-density environments are more prone to
mergers \citep{2001PAG} therefore other quenching processes like
harassment could also play a role.
Another process that can be at play in these environments is
strangulation, as there is no availability of the fresh gas and the
halo gas of these galaxies gets stripped by heating up
\citep{1980ApJL}.
Thermal evaporation \citep{1977NaturC} may also be present along these
processes. 

Ram pressure is expected to depend upon various parameters related to
in-falling galaxy and environment.  
To identify the importance of ram pressure stripping as quenching
mechanism one needs analytical models where we may study the effects
of various parameters on the amount of gas that gets stripped
\citep{2006ApJH,2018MNRASZ}. 
The basic aim of this paper is to study the efficiency of ram pressure
to remove the gas from a galaxy passing through different environments
(cluster, group and galactic) and its variation with galaxy and ambient medium 
parameters. \citet{koppen2018} have recently given 
an analytical treatment for ram pressure striping based on pure
kinematics of the system. However, they restrict there analysis to 
only cluster environment and a detailed dependence of striping on the 
system parameters is missing. \citet{2016Luo} have also studied the 
striping of cold gas in satellite galaxies only in a cluster, though their 
model doesnot alows for gas fall back or accretion.

A simplistic approach of a face-on disc 
galaxy passing through a spherically symmetric medium is considered here. 
The properties of the environment and the galaxy passing through it
will be discussed in details in the $\S$~\ref{sec:En} and
$\S$~\ref{sec:Gal}\,.
We begin with a galaxy modelled as a thin axisymmetric disc, passing
through a spherically symmetric cloud of gas.
In the first part we assume the galaxy to have a radial trajectory and
the infalling velocity profile, `$V(R_c)$' is explained in the
$\S$~\ref{sec:vel}\,.  
Non-radial trajectories for the infalling galaxies are also
considered, details of which are discussed in $\S$~\ref{sec:vel}\,.
We discuss the results of our numerical solutions for the model 
in $\S$~\ref{sec:results} for both radially infalling galaxies
($\S$~\ref{sec:radial}) and non-radial profiles for infalling galaxies
($\S$~\ref{sec:non-radial}). In $\S$~\ref{sec:inhomo} we breifly consider 
the effect of inhomogenities in the infalling galaxy on the amount of gas 
stripped due to ram pressure. 
Conclusions and discussion is presented in $\S$~\ref{sec:dis}\,.

\begin{figure}
\begin{center}
\includegraphics[scale=0.6]{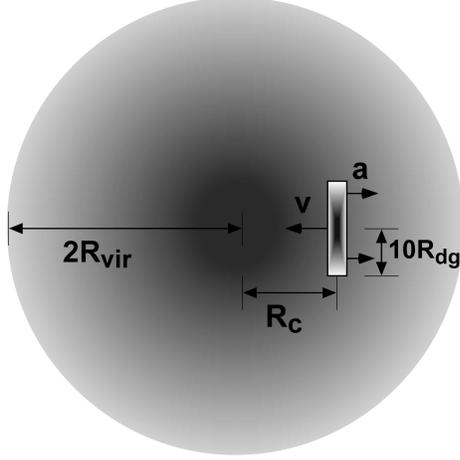}
\caption{Schematic of our model.
  The background or the ambient medium is a spherically symmetric
  sphere of radius $2 \times R_{\rm vir}$, where $R_{\rm vir}$ is the virial 
  radius as defined in Eq.~\ref{eq:r_virial}. The galaxy at radius $R_c$ in the 
  medium is moving with velocity $V$, entering the medium face-on. The 
  directions of velocity and acceleration, $a$ (defined in Eq.~\ref{eq:acceleration}) 
  are as indicated in the figure. $R_{\rm dg}$ is the disc scale length of the galaxy (Eq.~\ref{eq:rdg})}\label{fig:sch}  
\end{center}
\end{figure}

\section{Our Model}
\label{sec:model}

A schematic of our model is shown in fig.~\ref{fig:sch}, which we 
describe here.
The galactic disc is divided in annuli of equal width and the
force acting on each annulus while passing through the medium due to
Ram pressure is calculated at each instant.
Prescription for Ram pressure used in this paper is similar to what is
used by \citet{GG72}:
\begin{equation}
    F_{\rm ram}(R_c) = \rho(R_c) V^{2}(R_c), \label{F:ram}
\end{equation}
where $\rho(R_c)$ is the matter density of the medium from which the
galaxy is passing and $V(R_c)$ is the velocity of the galaxy relative
to the medium, both measured as a function of the radius of the
cloud.
This force acts on the gas component of a given annulus at radius $R$,
of the razor-thin disc, pushing it out.
Ram pressure is countered by a restoring force per unit area due to
the stellar component of the galaxy given by: 
\begin{equation}
    F_{\rm res}(R) = 2\pi G \Sigma_{\rm g}{}(R) \Sigma_{\rm s}{}(R), \label{F:res}
\end{equation}
where $\Sigma_{\rm g}$ and $\Sigma_{\rm s}$ are the surface densities
of gaseous and stellar components of the disc, respectively as defined in 
Sect.~\ref{sec:Gal} where we give the properties of infalling galaxy.  
$G$ is the gravitational constant.
Thus the acceleration experienced by the gas per unit area is given
by: 
\begin{equation}
a(R_c,R) = \frac{F_{\rm ram}(R_c) - F_{\rm res}(R)}{\Sigma_{\rm g}(R)}\label{eq:acceleration}.
\end{equation}
The acceleration on the disc annuli acts in the opposite direction
w.r.t. the velocity of the galactic disc as shown in
fig~\ref{fig:sch}\,.
If ram pressure dominates then the acceleration is positive, the gas
component of the annulus is pushed out of the galactic plane.
If a gas annulus has travelled more than one vertical scale height
($z_d$) above the disc plane, then it is considered as removed from
the disc. Gas once removed from the disc is removed from the system. 
Though our model 
allows for calculating the fall back of gas, the case of gas accretion 
to the galaxy, this happens when acceleration becomes negative. 
We donot calculate it 
here because once the stripped gas travels a distance more than $z_d$ away from 
the disc it is sufficiently outside the galaxies gravitational pull to be 
accreted by the galaxy\footnote{The only exception to this is an
  annulus that is stripped near a turning point in the trajectory of
  the galaxy through the medium.}. 
As the galaxy passes through the medium, the gas begins to strip, and
the radius up to which the gas is stripped at a given $R_{c}$ is noted
and is called $R_{\rm strip}$ (which is a function of $R_c$).
\begin{table*}
    \begin{tabular}{|c|c|c|c|} 
        \hline
        Part I: Ambient medium&&&\\
        \hline
        \hline
        Environment $\downarrow$ & $M$ (in $M_{\odot}$) & $z$ & $\delta$ \\
        \hline
        Cluster &$2 \times 10^{15}$ & $0, 0.5, 1$ & $1.0$ \\
        \hline
        Group & $2 \times 10^{13.5}$ & $0, 0.5, 1$& $1.0, 0.5$\\
        \hline
        Galactic &$10^{12}$ & $0, 0.5, 1$& $0.1, 0.3$\\
        \hline
        Part II: Infalling galaxy&&&\\
        \hline
        \hline
        Environment $\downarrow$&$M_{\rm gal}$ (in $M_{\odot}$)&$lambda$&$V_{\rm in}/V_{\rm vir}$\\
        \hline
        Cluster&$10^9 \textrm{to} 10^{13}$&$0.02$, $0.04$, $0.08$&$0$, $0.5$, $1.0$\\
        \hline
        Group&$10^6 \textrm{to} 10^{12}$&$0.02$, $0.04$, $0.08$&$0$, $0.5$, $1.0$\\
        \hline
        Galaxy&$10^6 \textrm{to} 10^{10}$&$0.02$, $0.04$, $0.08$&$0$, $0.5$, $1.0$\\
        \hline
    \end{tabular}
    \caption{The table contains two parts. Top one are the values of parameters used in the 
        study on which the background environment depends in all the three cases: Cluster, Group and Galactic. 
        The second part contains the values the model parameters on which the infalling galaxy depends and there 
        values used in different environments.}
    \label{tab:para_tab}
\end{table*}

\subsection{Environment}
\label{sec:En}

This section contains a brief account of the properties of the
background or the environment through which the galaxies pass.
We consider three classes of the environment in the present study:
cluster, group and galactic.
The background is modelled as a spherically symmetric cloud of gas.
The radial profiles of dark matter and baryons are assumed to be the
same for all the three cases, but with different total mass.
The mass, $M$ used for the cluster, group and galactic environment are
$10^{15} M_{\odot}$, $2 \times 10^{13.5} M_{\odot}$ and $10^{12}
M_{\odot}$, respectively.
Dark matter follows the NFW profile \citep{NFW}:
\begin{equation}\label{eq:nfw}
  \rho^{}_{\rm \scriptscriptstyle NFW}(R_c) = \frac{\delta_{c}
    \rho_{c}}{\left({R_c}/{R_{s}} \right) \left(1 + {R_c}/{R_{s}}
    \right)^{2}},  
\end{equation}
and
\begin{equation}
  \delta_{c} = \frac{200}{3} \frac{c^{3}}{ln(1+c)-
    {c}/{(1+c)}}, \label{eq:delc} 
\end{equation}
$\rho_{c}$ is the critical density of the Universe at the time when
the halo forms and is given by $\rho_c = 3H^{2}(z)/8\pi G$, and $R_{s}
= R_{\rm vir}/c$ is the scale factor.
The factor $H(z)$ is calculated using the Friedmann equations, see,
e.g., \citet{peebles93}. We use $\Lambda CDM$ cosmology with
$\Omega_{k} = 0.0$, $\Omega_\Lambda = 0.6911$, $\Omega_{m} = 0.3089$,
$H_{0}=67.8$ \citep{2016Plank}. 
The virial radius $R_{\rm vir}$ and the concentration parameter '$c$' 
for a virialized halo \citep{conref} of mass $M$ are calculated using;
\begin{align}
R_{\rm vir} = 0.784 &\left(\frac{M}{10^{8} h^{-1} M_{\odot}}\right)^{1/3}
\left[\frac{\Omega_m^z 18 \pi^2}{\Omega_m \Delta_c}
                      \right]^{1/3}\left(\frac{1 + z}{10
                      h^{-1}}\right)^{-1} kpc,\label{eq:r_virial}\\ 
c =&
     \frac{6.71}{\,\,\,\,(1+z)^{0.44}}\left(\frac{M}{2\times10^{12}h^{-1}M_{\odot}}\right)^{-0.092},\\   
\Omega_m^z =& \frac{\Omega_m (1 + z)^3}{\Omega_m(1 + z)^3 + \Omega_k(
              1 + z)^2 + \Omega_{\Lambda}},  
\end{align}
where $\Delta_c = 18.0\pi^2 + 82d - 39d^2$, $d = \Omega_m^z - 1.0$ and 
`$z$' is the redshift of the collapse which is assumed to same for
both ambient medium and infalling galaxy.  

The radial distribution of gas inside the dark matter halo are assumed
to follow Beta profile \citep{betaref}: 
\begin{equation}\label{eq:beta}
\rho_{\beta}(R_c) = \frac{\rho_{\rm g0}}{\left[1 + \left({R_c}/{R_{\rm
          core}}\right)^{2} \right]^{3\beta/2}}, 
\end{equation}
where $R_{\rm core}$ is the core radius which we take to be one-tenth
of the virial radius and $\rho_{\rm g0}$ is the central density. 
We take $\rho(R_c)$ (density of the ambient medium used in
eqn.~(\ref{F:ram})) to be equal to $\rho_{\beta}(R_c)$. Beta model serves 
as a test case, it being centrally concentrated is a resonable 
approximation \citep{1976A&AC,1986RvMPS}.  
To calculate the value of central density, we find the ratio of the total
gas mass gas ($M_{\rm g}$, say) and total dark matter mass ($M_{\rm
  bg}$, say) inside $R_{\rm vir}$ which is expressed as: 
\begin{equation}\label{eq:delta_def}
\Delta = M_{\rm g}/M_{\rm bg} = f_{uni} \delta,
\end{equation}
where $f_{uni}$ is the universal mass fraction taken to be $0.158$ 
\citep{2016Plank}, 
and $\delta$ is a parameter in our analysis, the value of which
depends upon the environment.
This parameter encodes the fraction of baryons retained in the form of
hot gas.
The value of $\beta$ for eqn.~(\ref{eq:beta}) is taken to be equal to
0.6 \citep{OM02}.
$\delta$ varies from 1.0 for the case of cluster, 1.0 \& 0.5 for
groups to, 0.3 \& 0.1 for the case of galaxy environment.
We have tended to err on the higher side for choosing values of
$\delta$ to simulate the worst case scenario \citep{psharma2012}. 
For a given value of $\Delta$ we solve eq~(\ref{eq:delta_def}) 
to get the value of $\rho_{\rm g0}$. In the first part of
table~\ref{tab:para_tab} we tabulate values of various parameters
pertaining to the ambient medium used for the present work.

\subsection{Infalling Galaxy}
\label{sec:Gal}

We discuss the geometry and properties of the infalling galaxies in
this section. 
We assume the galaxy to be a thin axisymmetric disc entering face-on
in the ambient medium. Though all galaxies are face-on in some projection, but the force being a vector quantity acts differently if the projections are taken 
into account. If the inclination is larger the effect on the gas is expected 
to be compresion in the front and striping from edges on the side. Here 
stripping mechanisms lead to different effects and needs to modelled
differently. 
Our analysis is applicable for inclinations upto $30$\textdegree\, in which case 
the force on the face-on side dominates than the edge-on. This also leads to 
enhanced effect of RPS in our model. 
Mass of the galaxy (baryons + dark matter) is $M_{\rm gal}$ and the
mass residing in the disc ($M_d$) is taken as: 
\begin{equation}
M_{\rm d} = f_{\rm d}\, f_{\rm uni} \, M_{\rm gal}.
\end{equation}
where $f_{\rm d}$ is the gas mass fraction in the galaxy disc and is a
parameter in the present work.
Two values of $f_{\rm d}$ that we consider are 0.1 and 0.3\,\citep{psharma2012}.
Out of this mass $M_{\rm d}$, $10\%$ is assumed to be in gas and rest
is in stars (these numbers are taken considering the Milky Way as a model, 
\citet{bt08}).
Radial profile of mass surface density of the galactic disc (baryonic)
is assumed to be an exponential with different scale lengths for the
gas and stellar components and is given by: 
\begin{equation}
\Sigma_{\rm d}(R) = \Sigma_{0} exp(-R/R_{\rm d}),
\end{equation}
where $R_{\rm d}$ is the scale radius of the disc.
In the following the scale radius for the gaseous component are
referred to as $R_{\rm dg}$ and $R_{\rm ds}$ for the case of stellar
disc. 
The corresponding surface densities are written as $\Sigma_{\rm dg}$
and $\Sigma_{\rm ds}$, respectively. 

We assume an isothermal profile for the dark matter halo of the
infalling galaxy, for ease of computation. The disc scale radius of the
galaxy in this case is given by:    
\begin{align}\label{eq:rdg}
R_{\rm dg} = \frac{\lambda}{\sqrt{2}} 0.784 & \left(\frac{M_{\rm
      gal}}{10^{8} h^{-1} M_{\odot}}\right)^{1/3} 
\left[\frac{\Omega_m^z 18.0\pi^2}{\Omega_m \Delta_c}
\right]^{1/3}\times\nonumber \\
& \times \left( \frac{1 + z}{10 h^{-1}}\right)^{-1} kpc,
\end{align}
where $\lambda$ is the spin parameter\footnote{The spin parameter is a
  dimensionless representation of the angular momentum of the Halo and
  is given by $\mathrm{\lambda = J|E|^{1 / 2} G^{-1} M^{-5 / 2}}$ where $E$ is the total energy of the halo, $M$ is the total mass of the halo, $J$ is the total angular momentum of the halo and $G$ is the gravitational constant.}
of the dark matter halo and the 
baryonic matter is assumed to share the same spin parameter.
$\lambda$ is also a parameter for the present study and it is varied
from $0.02-0.08$. 
The radial extent of galactic disc, $R_{\rm out}$ is taken to  be $10
\times R_{\rm dg}$.   
The stellar disc scale radius is obtained by \citep{bt08}:  
\begin{equation}\label{eq:rds}
R_{\rm ds} = \frac{1}{2}R_{\rm dg}.
\end{equation}

Vertical scale height, $z_{\rm d}$ of the disc is taken to be
$300$~pc, which is the same as that for the Galaxy \citep{bt08}.
It is kept at the same value for the galaxies of different masses
throughout the present work.
However, the present formulation is such that the conclusions are not
expected to change if we take scale height to be a function of mass,
as $z_{\rm d} = 300$~pc is the number for approximately the most 
massive galaxy used in our analysis and $z_{\rm d}$ is expected to 
decrease slowly for lower mass galaxies. Hence this gives a reasonable
estimate for $R_{\rm strip}$ for all masses of infalling galaxies.
To calculate $\Sigma_0$ for both gas and stellar disc we integrate
their respective surface density profiles from zero to $R_{\rm out}$.
This gives the total mass inside $R_{\rm out}$ and putting it equal to
their respective masses we find the value of $\Sigma_0$ for both gas
and stellar disc.  
These are referred to as $\Sigma_{0g}$ and $\Sigma_{0s}$, respectively.

\begin{figure*}
  \centering
  \subfigure[]{\includegraphics[scale=0.4]{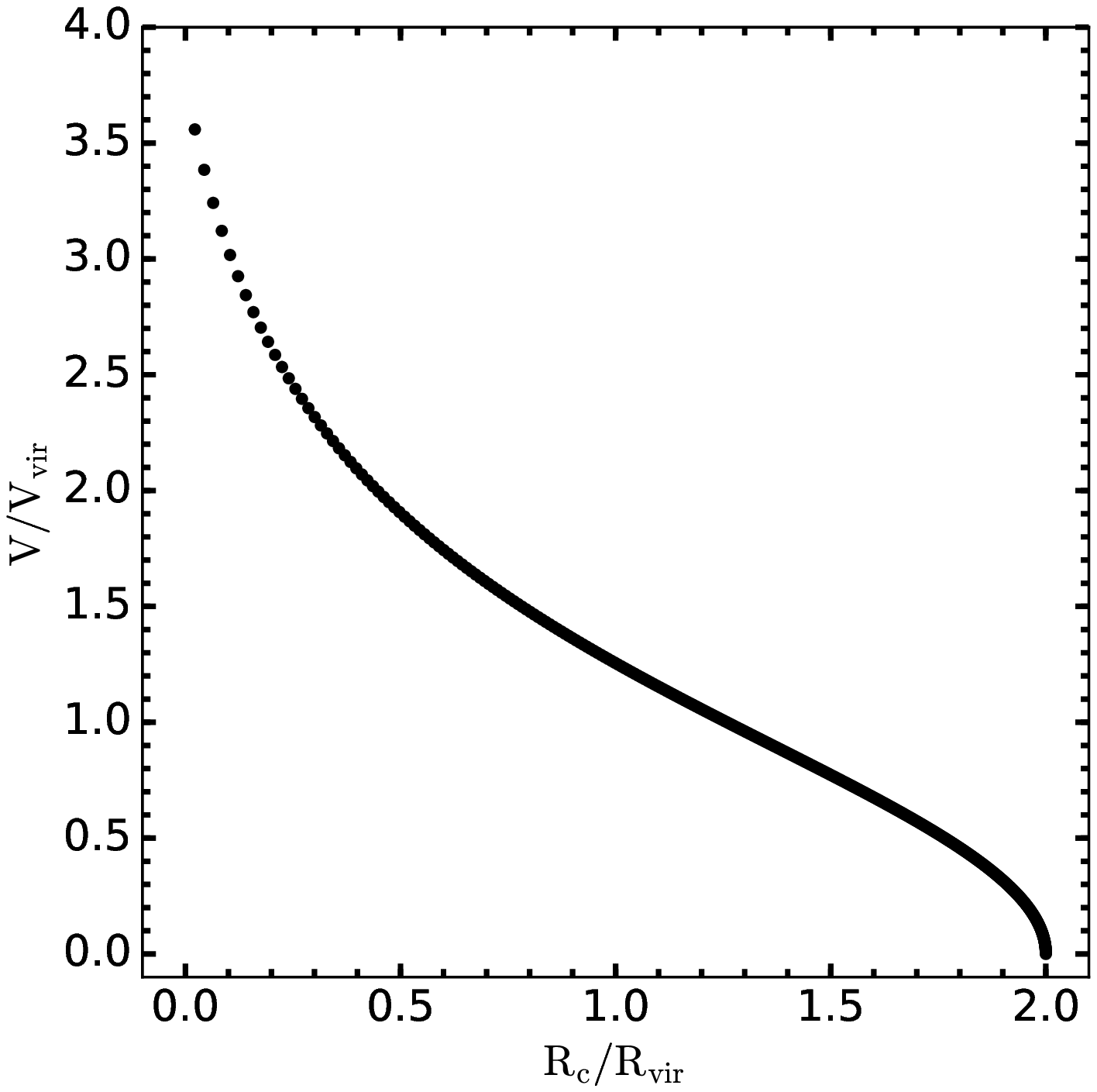}}\quad
  \subfigure[]{\includegraphics[scale=0.4]{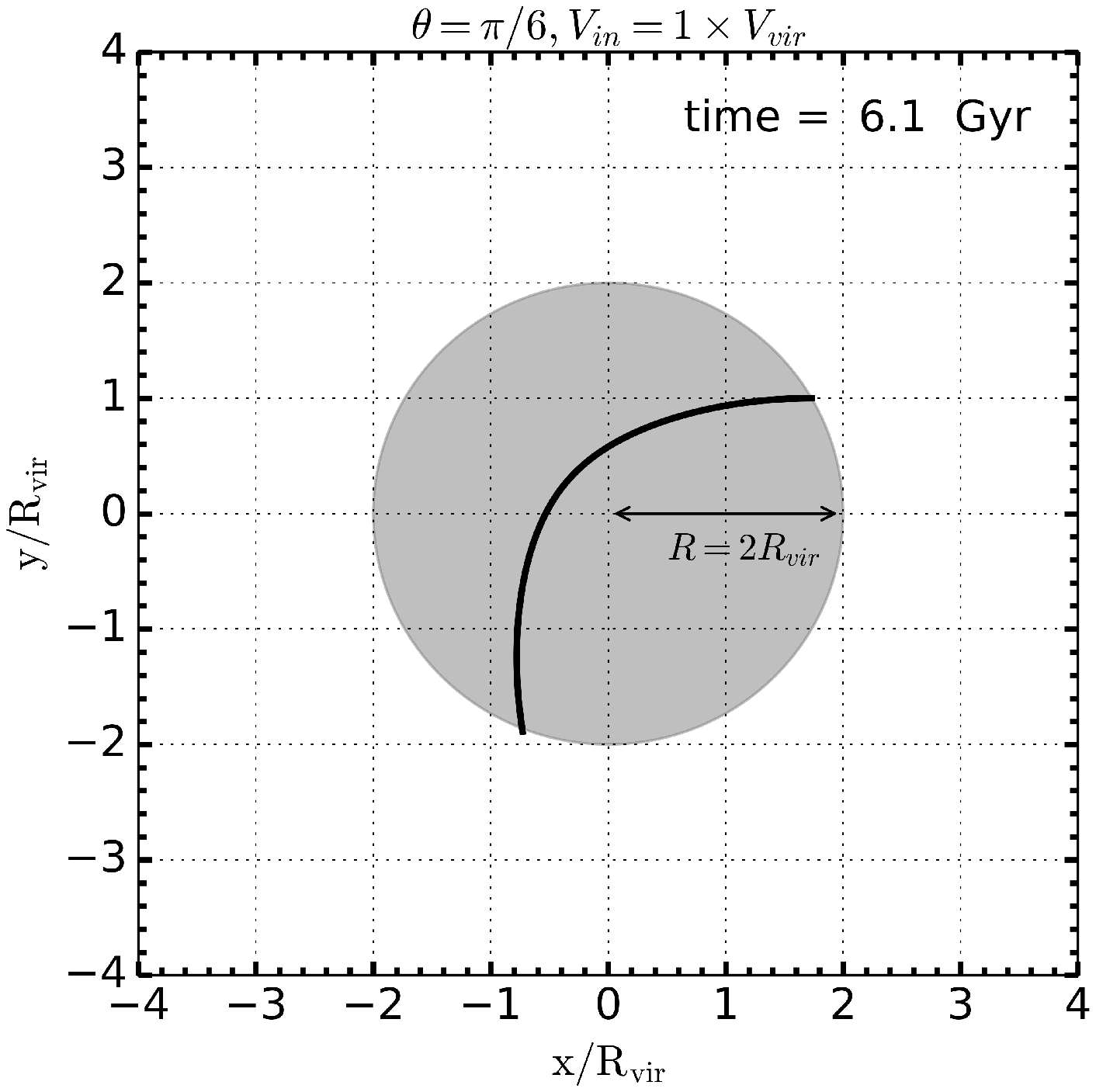}}\label{fig:traj}
  \caption{(a)Variation of velocity with $R_c$. The galaxy starts at
    zero velocity at $2 \times R_{\rm vir}$ which increases as it passes 
    through the ambient medium. (b) The plot for the trajectory of an
    infalling galaxy for non-radial infall in the $x-y$ plane. The
    axis is normalised w.r.t. $R_{\rm vir}$ and all other parameters
    are given in the plot.}\label{fig2}
\end{figure*}

\subsection{Velocity Profile}\label{sec:vel}

The infalling galaxy is first taken to be on a radial trajectory,
entering face-on in the medium.
The galaxy starts at rest at $2R_{\rm vir}$, and it accelerates during
infall into the halo.
The infalling velocity profile is shown in fig~\ref{fig2}(a).
The $x-$axis is the ratio of $R_c/R_{\rm vir}$ and the $y-$axis is
$V(R_c)/V_{\rm vir}$, where $V_{\rm vir}$ is the circular velocity at
the virial radius and is given by 
\begin{equation}
V_{\rm vir} = \left(\frac{M}{10^{8} h^{-1} M_{\odot}}\right)^{1/3}
\left[\frac{\Omega_m^z 18\pi^2}{\Omega_m \Delta_c}
\right]^{1/6}\left(\frac{1 + z}{10 h^{-1}}\right)^{1/2} km s^{-1}. 
\end{equation}
Where all the terms have their usual meanings and are discussed earlier 
in the text. 
We also vary the initial velocity $V_{\rm in}$ (i.e. the velocity of
the infalling galaxy at $2R_{\rm vir}$) from zero to $V_{\rm vir}$.
We have taken the galaxy to go all the way up to the centre of the
halo along a radial orbit.
This has been done to study the worst case scenario.
In a realistic scenario, galaxies enter halos on different types of
orbits, and a large fraction of galaxies do not approach the centre in
the first crossing.   
The range of all the parameters of the infalling galaxy is given
in part-II of table~\ref{tab:para_tab}. 

We also study the scenario where the infalling galaxy has a finite
angular momentum and hence does not go to the centre of the halo.
This facilitates comparison with observations even though in all other
aspects we continue to work with the worst case scenario.
The galaxy enters the cluster with a finite impact parameter $b$
w.r.t. the centre of the central halo.
The impact parameter is converted to an angle ($\theta$, say)
subtended at the centre, which is a parameter in our analysis.
The radial infall is $\theta =0$, which corresponds to $b = 0$.
As we increase $b$, $\theta$ can be calculated as $\sin(\theta) =
b/2R_{\rm vir}$.
The maximum value of $\theta$ is taken at $b = R_{\rm vir}$, and we
get $\theta_{\rm max} = \pi/6$\,.
An example of the trajectory, in the $x-y$ plane, which the infalling
galaxy follow is shown in figure~\ref{fig2}(b).  

\begin{figure}
\begin{center}
\includegraphics[height=8cm,width=9cm]{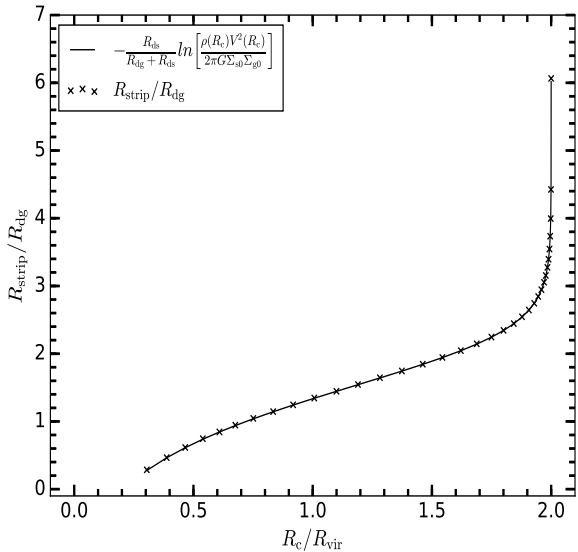}
\caption{Stripping radius versus radius of the ambient medium for
  cluster environment at $z = 0$, for $M_{\rm gal} =
  10^{12}M_{\odot}$, $\lambda = 0.04$, $f_{\rm d} = 0.3$ and $\delta =
  1.0$. The solid line is calculated using eqn.~(\ref{eq:r_strip}) and
  the points are calculated using our model.}\label{fig:comp} 
\end{center}
\end{figure}

\section{Results}
\label{sec:results}

This section contains the results obtained by numerically solving the
model proposed in the above section.
We have worked with {\em ASTROPY} \citep{astropy18}, which is a
a package that allows working in dimensional units.

The quantity we are interested in, and we calculate, is the radius
up to which the gas in the galactic disc is stripped, $R_{\rm strip}$
as the galaxy passes through the ambient medium. In all the results 
from here on we shall normalise 
the stripping radius with $R_{\rm dg}$ and $R_c$ with the virial radius.
Striping radius directly translates to the fraction of gas mass removed
from the disc as: 
\begin{equation}
  M_{\rm removed} = \frac{1}{0.1M_{\rm d}}\intgbth{R_{\rm
      strip}}{R_{\rm out}} 2\pi x \Sigma_{\rm dg}(x)\dmath x. 
  \label{eq:m_removed}
\end{equation}
To validate the numerical scheme used here we first reproduce the
results of \citet{GG72}. 
We also get an order of magnitude estimate of the stripping radius as
a function of $R_c$, which can be done by solving the equation $a(R_c, R_{\rm
  strip}) = 0$ to get; 
\begin{equation}
\frac{R_{\rm strip}}{R_{\rm dg}} = -\frac{R_{\rm ds}}{R_{\rm dg} + R_{\rm ds}}
\ln\left[\frac{\rho(R_c)V^{2}(R_c)}{2\pi
    G\Sigma_{0g}\Sigma_{0s}}\right]\label{eq:r_strip} 
\end{equation}
In fig.~\ref{fig:comp} we plot the stripping radius versus the radius
of ambient medium calculated using eqn.~(\ref{eq:r_strip}) for cluster
environment at $z = 0$, for $M_{\rm gal} = 10^{12}M_{\odot}$, $\lambda
= 0.04$, $f_d = 0.3$ and $\delta = 1.0$.
Crosses on the plot show the values calculated from our model, for the
same set of parameter values.
The match is quite satisfactory, validating our numerical scheme.

We divide our results into three sections based on their respective
environments and discuss the dependence of $R_{\rm strip}$ on various
parameters separately.
We further divide our results into two parts, the first one is for
radially infalling galaxies and the second one is for the galaxies
falling with some impact parameter w.r.t. the central galaxies in
clusters.

\begin{figure*}
\begin{center}
\includegraphics[scale=0.45]{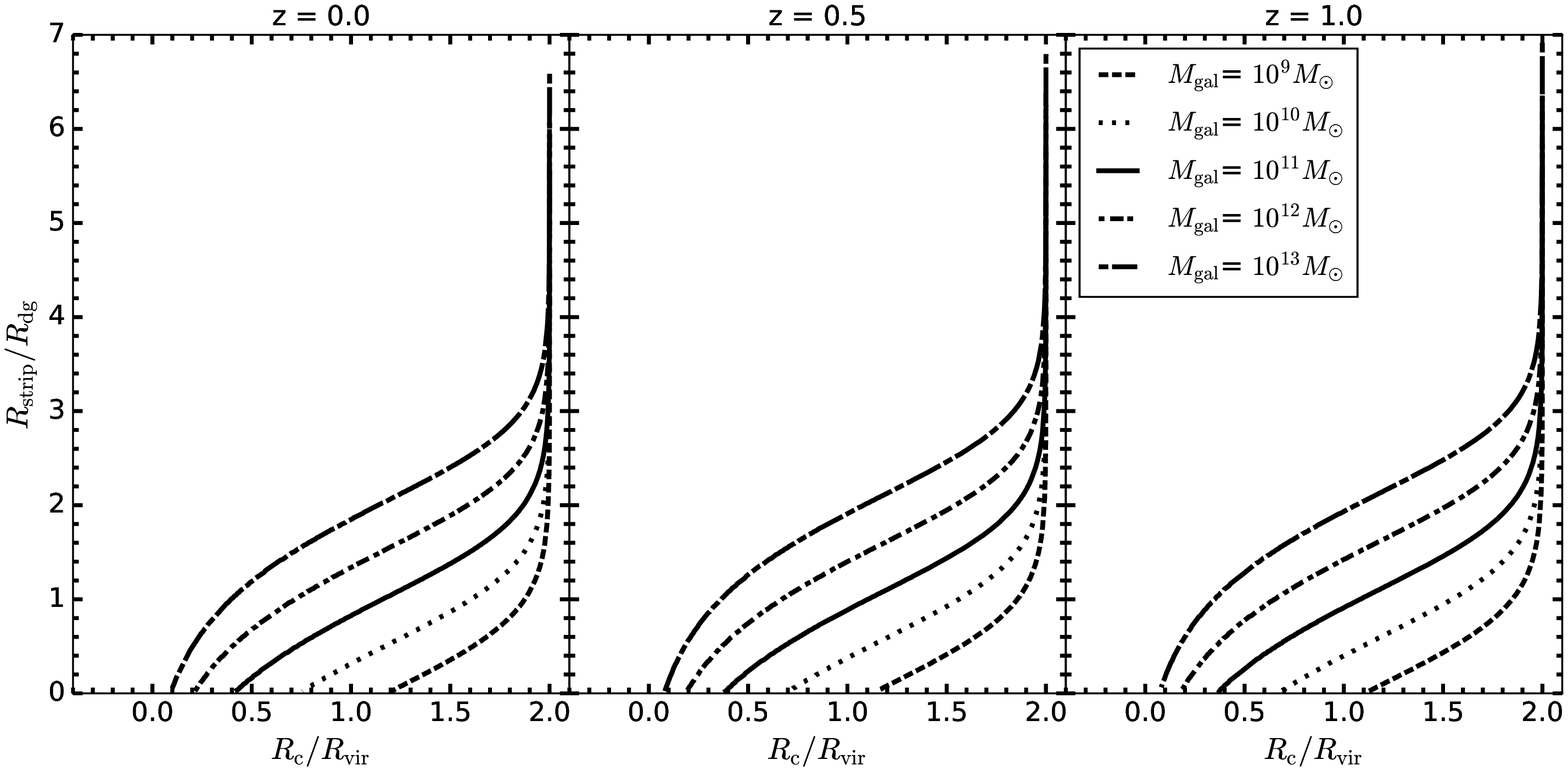}\\
\includegraphics[scale=0.45]{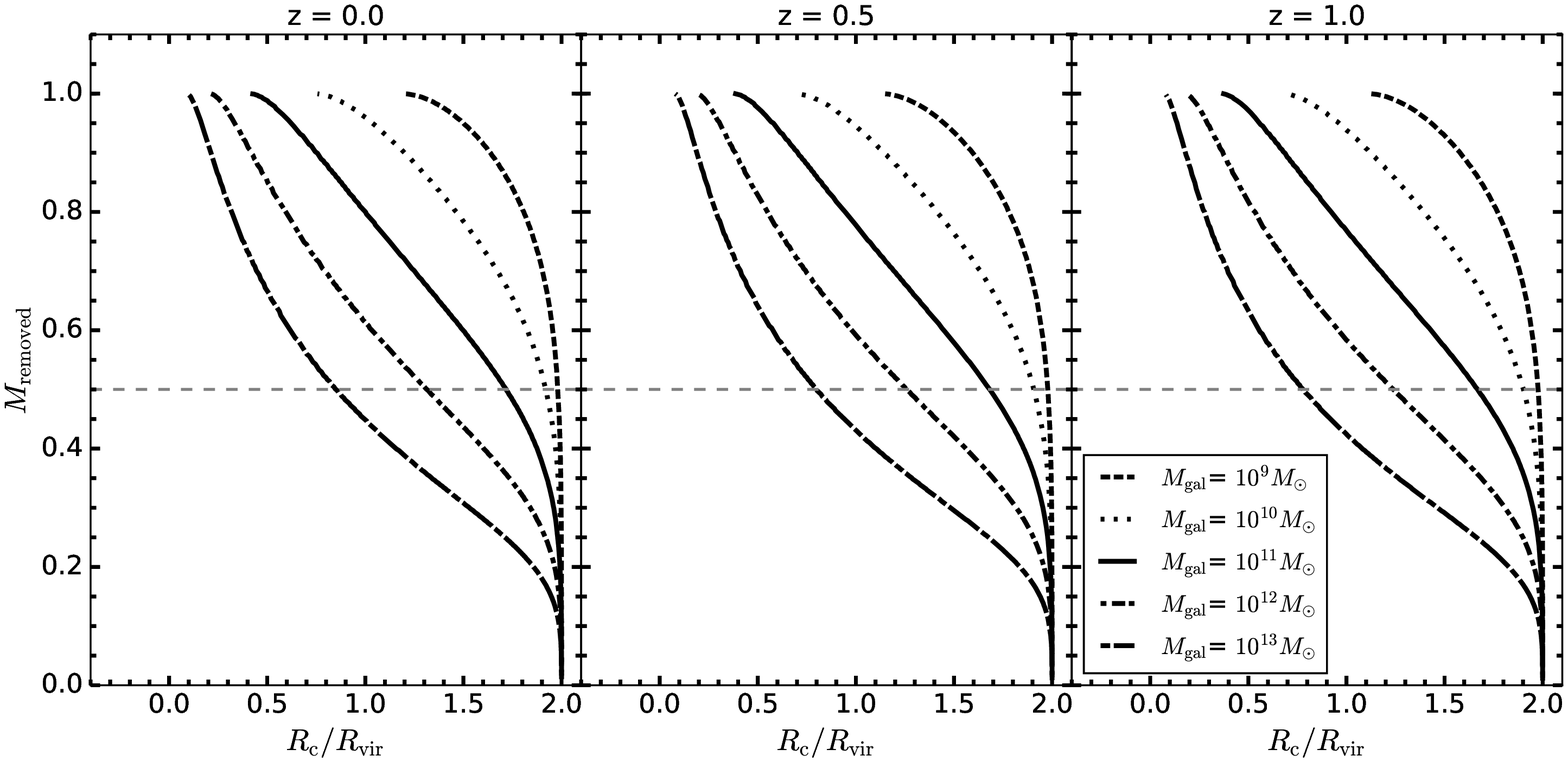}
\caption{Top panel is the plot of striping radius versus cluster
  radius for the same parameters as that of fig.~\ref{fig:comp}, for
  different values of $z$ and $M_{\rm gal}$.  Lower panel is the plot
  of $M_{\rm removed}$ calculated using
  eqn.~(\ref{eq:m_removed}), with dashed horizontal line marking the position of 
  $M_{\rm removed} = 0.5$.}\label{fig:r_strip} 
\end{center}
\end{figure*}
\begin{figure*}
\begin{center}
\includegraphics[scale=0.45]{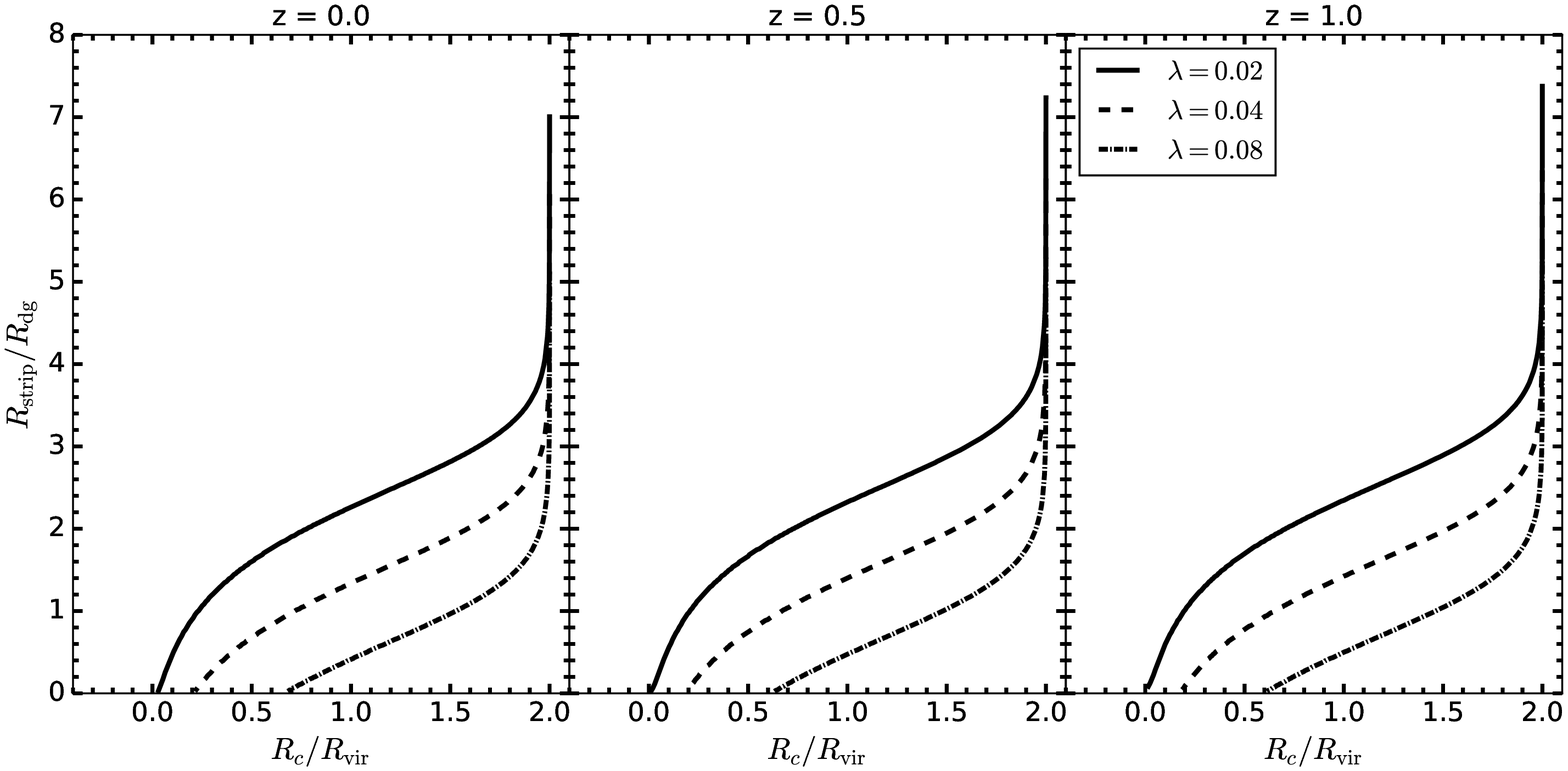}\\
\includegraphics[scale=0.45]{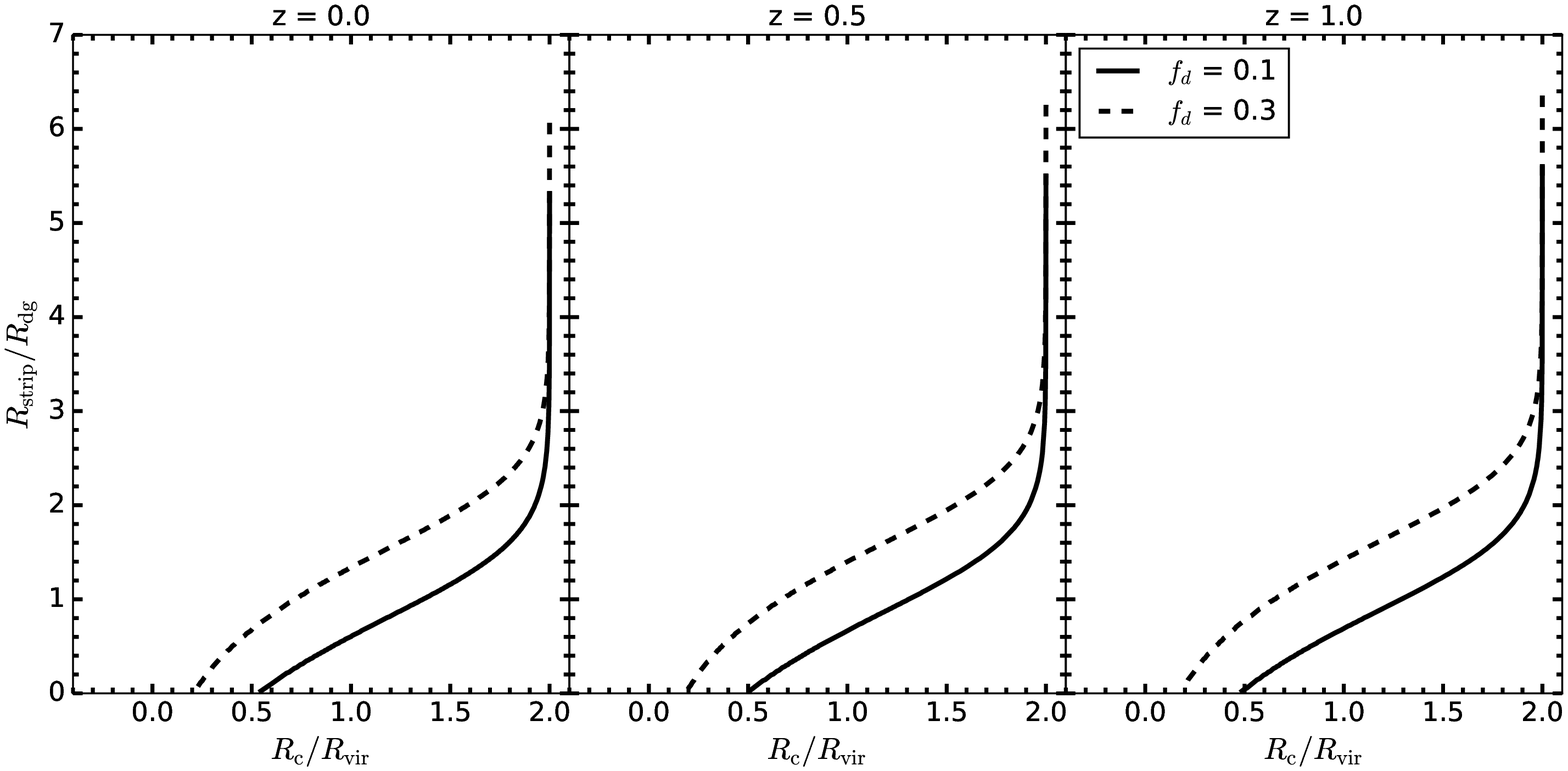}\\
\includegraphics[scale=0.45]{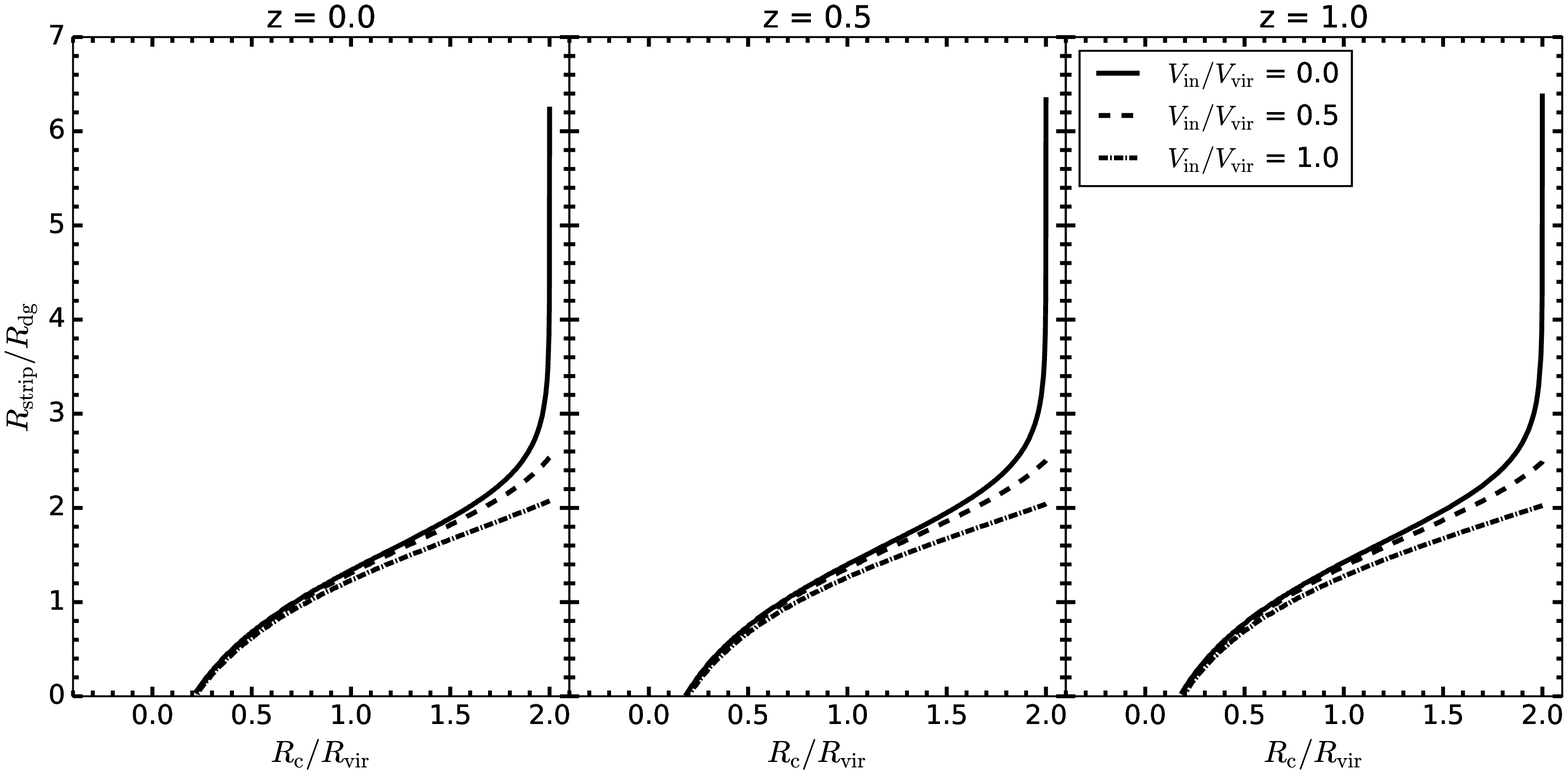}
\caption{Variation of stripping radius with spin parameter, $\lambda$
  (top panel), baryonic mass fraction, $f_{\rm d}$ (middle panel) and
  infalling velocity lower panel. The infalling galaxy is taken to be
  $10^{12} M_{\odot}$.}\label{fig:lambda} 
\end{center}
\end{figure*}

\subsection{Radial infall}\label{sec:radial}

Here we discuss the results for the case of radial infall of galaxies
for all the background environments. 

\subsubsection{Cluster Environment}

We start by studying the variation of stripping radius $R_{\rm
  strip}/R_{\rm dg}$ with mass of the infalling galaxy, $M_{\rm gal}$
in fig.~\ref{fig:r_strip}.
We show the results for different redshifts. The lower panel shows the mass stripped or removed due to ram pressure
stripping as the infalling galaxy comes closer to the centre. The
dashed horizontal line here gives  
the position where $50\%$ of gas is removed from he disc. 
Different line types refer to infalling galaxies of different mass. 
We have used $\lambda = 0.04$, $f_d = 0.3$ and $\delta = 1.0$.
We see that as the infalling galaxy goes closer to the centre of the
cluster more and more gas gets stripped, eventually striping the whole
gas disc.
The value of $R_c$ at which the disc gets stripped is related to the
mass of the infalling galaxy: lower mass galaxies lose the gas at
larger $R_c$.
It also follows that low mass galaxies get stripped in less time.
We note that there is no variation with redshift, as can be seen by
comparing the panels for $z=0$, $0.5$ and $1.0$.

This result may appear counter-intuitive as both $\rho_{\beta}(R_c)$
and $V(R_c)$ both increase with $z$ and hence the effect of ram 
pressure should be more prominent at higher redshift.
However, a closer look at eqn.~(\ref{eq:r_strip}) shows that the
$z$-variation in both numerator and denominator are same and get
cancels.
This leaves $R_{\rm strip}/R_{\rm dg}$ independent of the redshift,
which is what we observe in our numerical results.
This cancellation requires that the infalling galaxy and the central
halo form at the same redshift.
If the infalling galaxy forms at a higher redshift then the disc has a
higher surface density and can be expected to resist ram pressure
stripping.
We do not explore this variation as we are working with the worst case
scenario. We do not show the plots for $M_{\rm gal} < 10^9$ as the whole gas
disc in those cases gets stripped at the periphery of the clusters.
We shall discuss the implications of these observations in $\S~\ref{sec:dis}$. 

Other parameters that we explore in the present study are the spin
parameter $\lambda$, baryonic mass fraction of total mass residing in
the infalling galaxy $f_{\rm d}$ and the initial velocity of the
galaxy $V_{\rm in}$.  
In fig.~\ref{fig:lambda} we give the plots for how the stripping
radius changes with variation in these parameters. 
The results are shown for $M_{\rm gal} = 10^{12} M_{\odot}$.
The top panel shows the results for different values of $\lambda$, 
for all $z$-values for galaxy starting at rest and $f_{\rm d} = 0.3$.
The scaled radius of the infalling disc is directly proportional to
$\lambda$ as can be seen in eqn.~(\ref{eq:rdg}).
Hence, the larger the value of $\lambda$ larger is the radial extension of
the disc and smaller is the surface densities of both gas and stars
implying that the restoring force on the gas disc will be lesser,
making the ram pressure stripping more efficient.
This distinction is clearly seen in the top panel of
fig.~\ref{fig:lambda}. 
Note that we have assumed the same fraction for mass in gas and stars
in the disc for all the value of the spin parameter.
This may not be true, and this may affect the results. 
In particular if the gas fraction is higher in galaxies with a higher
spin parameter such that the product $\Sigma_{0g} \Sigma_{0s}$ takes on a
higher value for the same disc mass, then the stripping will be less
pronounced than in our model.

The middle panel is the variation with $f_{\rm d}$ at all the
redshifts.
For these plots, it is assumed that the galaxy starts with zero
initial velocity at twice the virial radius and we use $\lambda =
0.04$.
We observe that for higher values of $f_{\rm d}$ the gas disc can
sustain itself against ram pressure for more time.
The reason for this is that higher baryon density implies a higher surface 
density of the galactic disc, and hence higher is the
restoring force, allowing the disc to stay intact for a longer time. 
The bottom panel gives the variation with $V_{\rm in}$ for $\lambda =
0.04$ and $f_{\rm d} = 0.3$.
We see that as the galaxy enters the medium, the stripping is initially
stronger for larger initial velocity but eventually the stripping
radius becomes independent of the initial velocity and the radius at
which the whole gas disc is stripped is same for all the cases.
Thus variation in surface density and spin parameter results in a
stronger change in ram pressure stripping, whereas initial velocity
does not affect the results significantly.

For all the three parameters that we explore in fig.~\ref{fig:lambda},
we have also shown the variation with redshift. 
Again, we do not see any variation with redshift, in agreement with
arguments presented above.

\begin{figure}
\begin{center}
\includegraphics[height=6cm]{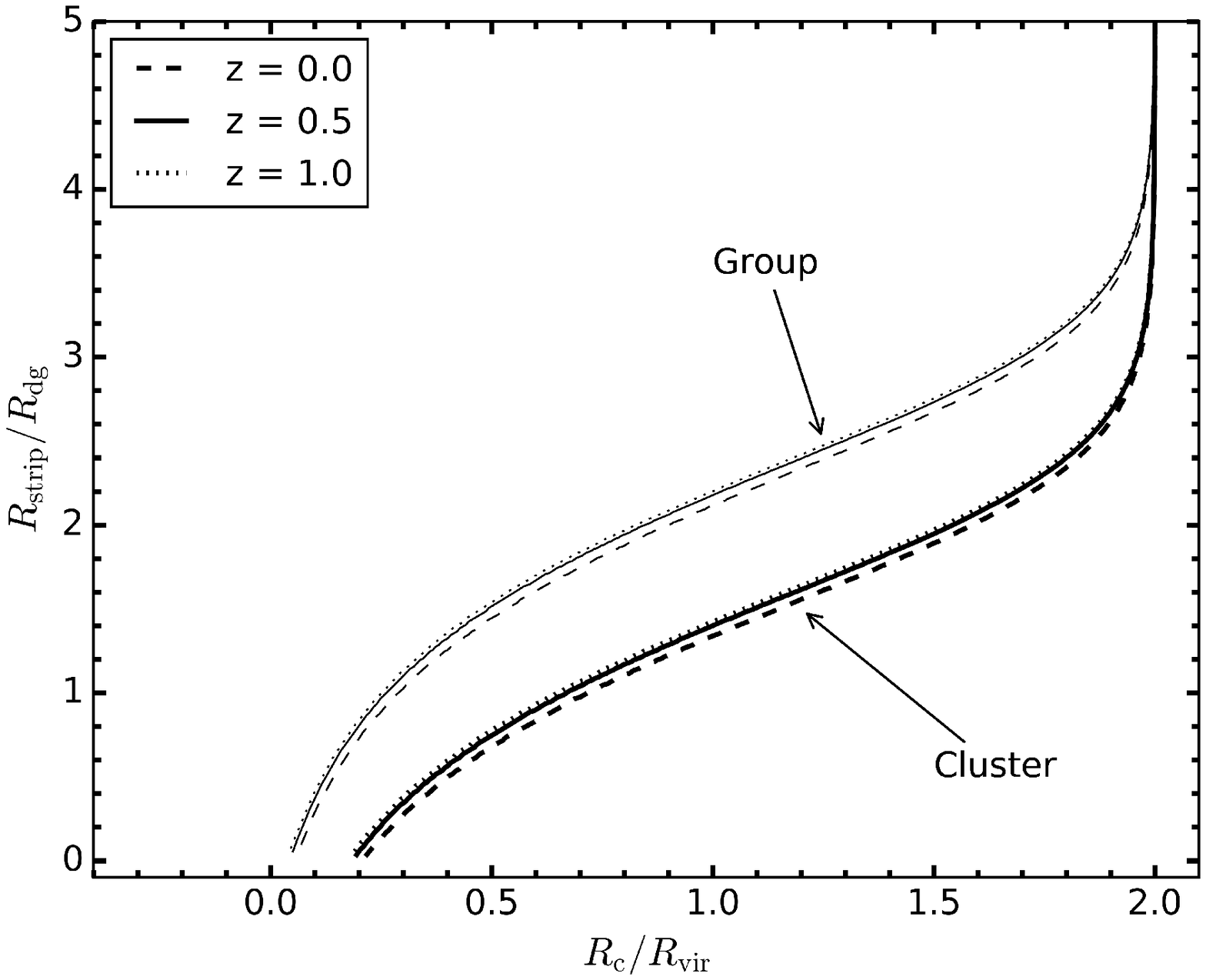}
\caption{The comparison for stripping radius for cluster (thick, lower curves) 
  and group (upper set of curves) environment for the infalling galaxy of same 
  mass ($10^{12}M_{\odot}$) and different redshifts.}\label{fig:group} 
\end{center}
\end{figure}

\subsubsection{Group Environment}

We next describe ram pressure stripping in a group environment where
the halo mass of the group is assumed to be $2 \times
10^{13.5}$~M$_{\odot}$. 
The density of the ambient medium decreases as we decrease the mass,
suppressing the effect of ram pressure.  
The same is verified by our numerical results which are plotted in
fig.~\ref{fig:group}.  
For this plot we have taken the galaxy of mass $10^{12} M_{\odot}$ to
enter the medium at rest, at a redshift of zero, $\delta = 1.0$ and
$\lambda = 0.04$\,.  
In this, we plot we show $R_{\rm strip}$ vs $R_{\rm c}$ for both
clusters and groups at different redshifts.
The thick, lower curves are for the cluster environment, and the upper
set of curves are for the group environment.  
Striping in groups is slower as compared to clusters and galaxies are
able to reach closer to the centre without being stripped of gas. 
The $z$-variation is negligible, as discussed above.

Other parameters such as $M_{\rm gal}$, $z$, $\lambda$ are also varied
with ranges given in table~\ref{tab:para_tab}.
The variation of stripping radius for all these follow the same 
trends as was the case for the cluster environment.
The stripping radius is shifted to higher values for groups as
compared to clusters, in consonance with the example shown in
fig.~\ref{fig:group}. 
To avoid repetitions, we do not plot them here. 

\begin{figure}
\begin{center}
\includegraphics[scale=0.4]{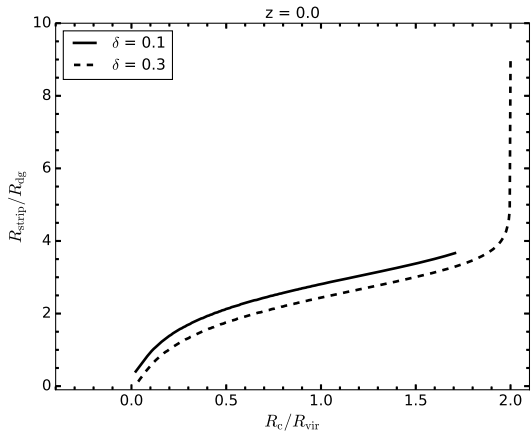}
\caption{The variation of $R_{\rm strip}$ for the case of galaxy
  environment for an infalling galaxy of mass $10^{10}
  M_{\odot}$.}\label{fig:galaxy1} 
\end{center}
\end{figure}

\begin{figure*}
  \begin{center}
    \includegraphics[scale=0.5]{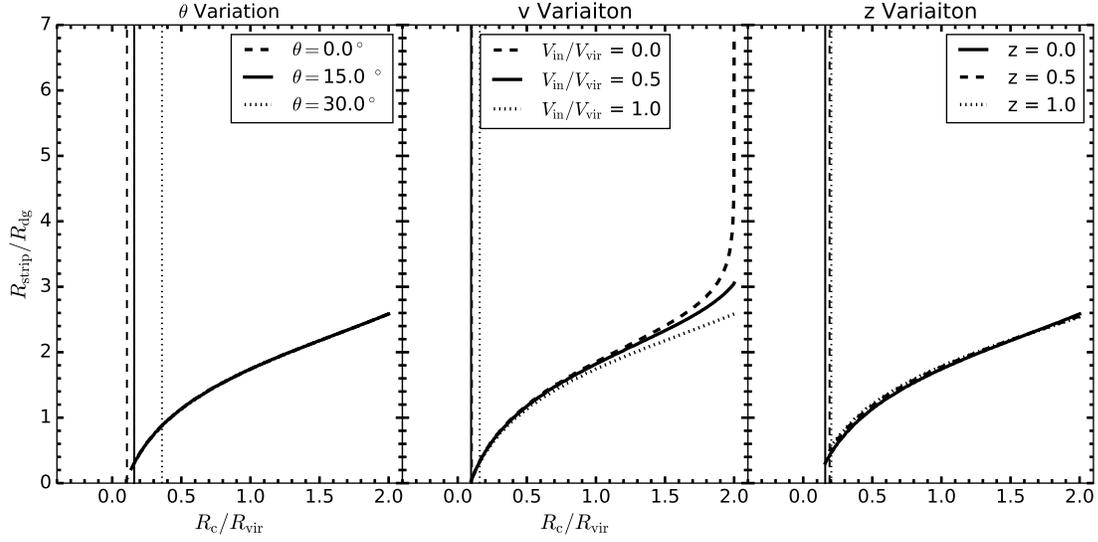}
    \caption{The plots of $R_{\rm strip}$ with $\theta$, $V_{\rm in}/V_{\rm
        virial}$, and $z$ for an infalling galaxy of mass $10^{13}
      M_{\odot}$ with $\lambda = 0.04$. The vertical line show the last 
      point in each curves. Second and the third panels are the plots 
      for $\theta = 15$\textdegree. }\label{fig:trajectory1} 
  \end{center}
\end{figure*}

\subsubsection{Galaxy Environment}

The third environment we study is the galaxy environment.
In this case, the density of the ambient medium is lower than that for
a group and cluster.
This and the lower velocity of infalling galaxy result in less
efficient ram pressure stripping.
The stripping radius for an infalling galaxy of mass $M_{\rm gal} =
10^{10} M_{\odot}$ with zero initial velocity and other parameters
same as that of the case of groups is plotted in
fig.~\ref{fig:galaxy1}.
Two different lines are for different values of $\delta$, i.e. gas
mass fraction in the ambient medium.
Gas in the ambient medium is what is causing ram pressure stripping.
Gas in the infalling galaxy strips faster for higher values of
$\delta$.  

\subsection{Non-radial infall}\label{sec:non-radial}

Galaxies fall into clusters on different types of orbits, with the
generic orbit being non-radial.  
It implies that most galaxies do not come very close to the centre of
cluster and hence RAM pressure stripping may be much less effective.
To study the effect of non-radial motion, we vary the trajectory of
infalling galaxies. 
The parameter that measures the deviation from radial infall is
$\theta$ as introduced in section-\ref{sec:vel}.
The variation of $R_{\rm strip}$ with $\theta$, $V_{\rm in}/V_{\rm
  virial}$, and $z$ are plotted in fig.~\ref{fig:trajectory1}.
Variation of stripping radius as different parameters are varied is
shown here.
The infalling galaxy is taken to have a mass of $10^{13}$~M$_{\odot}$
with $\lambda = 0.04$ for this figure. 

In the first panel, we show the variation with $\theta$, for $V_{\rm
  in}/V_{\rm virial} = 1$.
As the angular momentum of the infalling galaxy about the centre of
the cluster increases, stripping becomes less effective.
We observe that for $\theta = 30$\textdegree\, 
the gas in the disc is retained up to the scale radius $R_{\rm dg}$.
Thus non-radial infalling trajectory can lead to retention of some gas
in a galaxy. Second and the third panels are the plots for 
$\theta = 15$\textdegree\, .  
The variation with velocity and other parameters follows the same
trends as discussed for radially falling galaxy. 
\begin{figure*}
\begin{center}
	\includegraphics[scale=0.5]{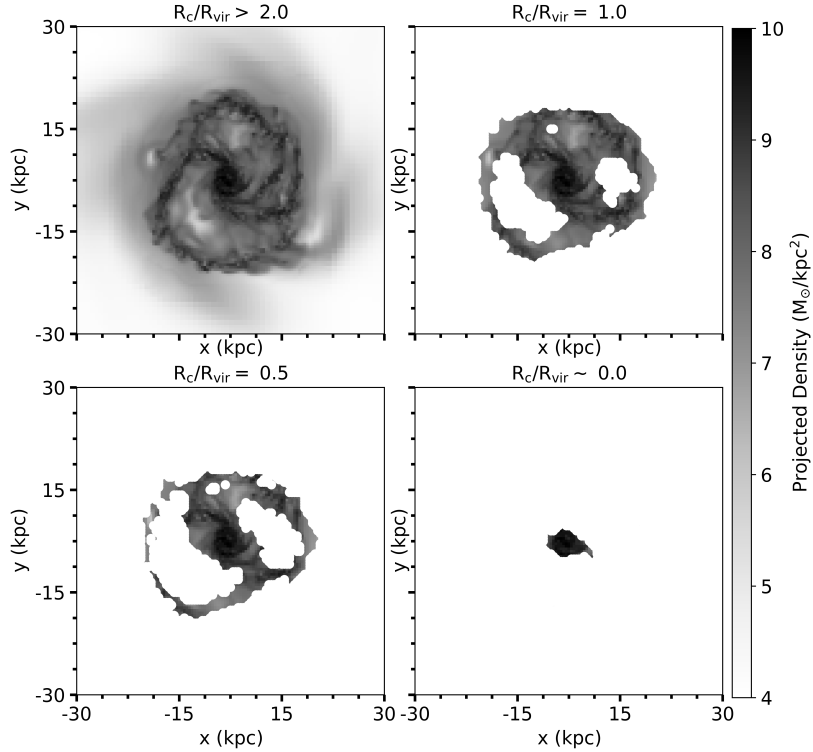}
	\caption{Infalling galaxy (as shown in the top left panel) with constant velocity of $\rm V_{vir}$ 
	is subjected to radially varying intra-cluster density. We plot here gas projection at different 
	cluster-centric distances, as given on the top of each panel.}
	\label{fig:strip}
	\end{center}
\end{figure*}
\begin{figure}
	\begin{center}
	\includegraphics[scale=0.5]{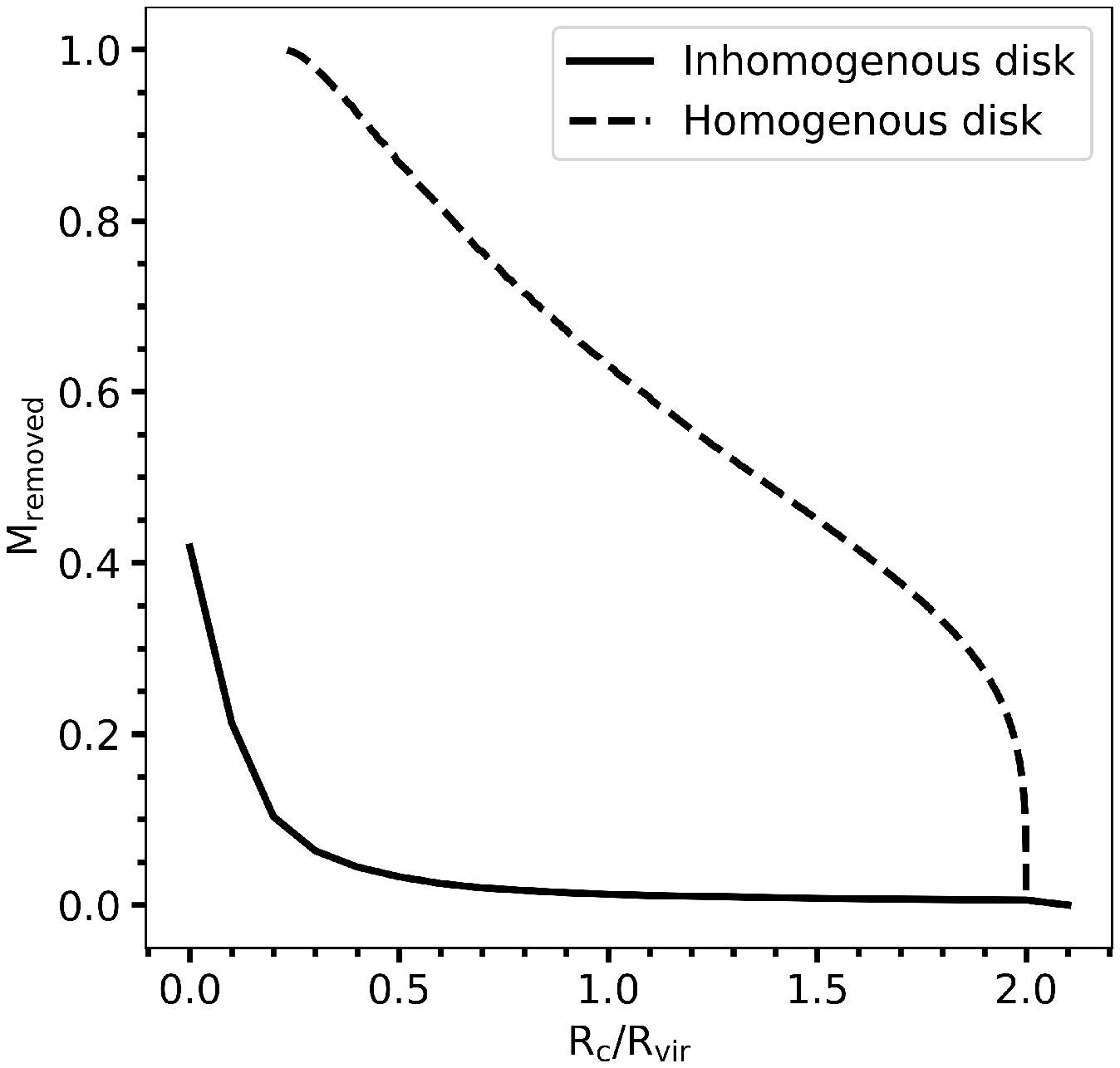}
	\caption{Fraction of mass removed as galaxy is subjected to ram pressure stripping 
	plotted against cluster centric distance ($R_{c}$) scaled with virial radius of cluster 
	($R_{\rm vir}$), for both, inhomogenous (solid line) and homogenous discs (dotted line).}
	\label{fig:mass_fraction}
	\end{center}
\end{figure}

\subsection{Inhomogenous infalling disc}\label{sec:inhomo}

In order to test our analytical model and explore sources of large
deviations from the worst case scenario, we study striping of an
inhomogeneous disc galaxy.
To simulate the infalling galaxy we use Adaptive Mesh Refinement (AMR)
code RAMSES (\cite{2002Teyssier}.
A box size of 300 kpc with maximum resolution of $\mathrm{\sim \ 36
  \ pc}$ is taken.
Simulated galaxy has setting similar to that of G$10$ galaxy from
\cite{2017Rosdahl}, with initial galaxy mass as $M_{200}$ of
$9\times10^{11}M_{\odot}$, with disc fraction of  $12\%$  and gas
fraction of $5\%$.
Stellar and gas disc both have exponential profile with disc scale
length of $3.2$ kpc.
We do not include any bulge in our simulation from initial condition.
We evolve the galaxy to get gas and stellar distribution at $450$ Myr.
Then a projection is taken to compute surface densities for gas and
stars. 
Top left of figure~\ref{fig:strip} shows the surface density. 
 
The gas and stellar surface density from this simulation is then used
to study the ram pressure stripping using our analytical model.  
We take a simple can of galaxy coming with $\rm V_{vir}$ subjected to
radially varying density of intra-cluster medium.
Panels of fig.~\ref{fig:strip} are the snap-shots from our model as
the galaxy passes through the ambient medium.
The RPS is less effective in the regions where the disc had 
enhanced density and there is sufficient amount of gas left in the 
disc of galaxy, even
as the galaxy passes through central region of the cluster.
Thus star formation can continue in galaxies falling into a cluster
halo for at least one crossing time as gas in the denser molecular
clouds is not striped off. 
To illustrate this further in fig.~\ref{fig:mass_fraction} we plot the
fraction of total mass removed versus $R_c/R_{\rm vir}$.
A litle more than $40\%$ of the gas mass gets removed in case of
inhomogeneous disc in contrast to $100\%$ for the case of previously
considered homogenous disc.
We thus conclude that though RPS is an effective mechanism of removing
the gas from the disc galaxies but the internal structure of the
infalling galaxy has a larger role to play. 
Thus inhomogeneities in the inter-stellar medium of galaxies play a
very significant role in retention of gas and hence in star formation
in spite of the effects of ram pressure. 

\section{Conclusions and Discussions}\label{sec:dis}

This paper is a study to answer the question that: {\it How effective
  is ram pressure stripping to remove gas from the galaxy as it passes
  through the ambient medium in a variety of halos and varying group of 
  infalling galaxies?} 
We consider the ambient medium to be spherically symmetric and the
disc galaxy enters the medium face-on.
We consider both radial and non-radial initial velocities for the
galaxies.
We study three kinds of environments namely; cluster, group and
galaxy.
These differ from each other in their mass and retention of hot gas in
the halo.
The different depth of the potential well translates into different
typical velocities with which galaxies fall in.
We assume an isothermal profile for halos and a beta model for the gas
distribution of ambient medium, the disc of the infalling galaxy is
assumed to have an exponential surface density profile. 
The ambient medium and infalling galaxy profiles depend on various
parameters: the redshift of formation, spin parameter, baryonic mass
fraction etc.
These have been discussed in details in $\S$~\ref{sec:model}.
We model discs of galaxies as a set of annuli, and we test for ram
pressure stripping for each annulus.
This allows us to compute the stripping radius, i.e., the radius
beyond which all gas from the disc has been stripped. 
Stripping radius is directly related to the fraction of gas mass
removed from the galaxy.
We compute the stripping radius in units of the gas disc scaling
radius.
This is pertinent for interpretation as much of star formation in disc
galaxies happen around the scale radius, extending to a few scale
radii at most. 
Thus, if stripping radius is of the order of gas disc scale radius or
smaller then we expect star formation to be quenched.
However, if the stripping radius is a few times the gas disc scale
radius or larger, then star formation in such a galaxy may not be
affected significantly. 

The results of our study have been enumerated in
section~\ref{sec:results} in detail.
Our approach throughout has been to consider the worst case scenario,
one where we maximum amount of gas is stripped from the galaxy.
We outline the processes that may lead to gas being retained in the
discussion below. 
Here we summarise our major findings.
\begin{itemize}
\item
  For infall in a cluster of galaxies, galaxies of all masses lose the 
  gas disc for radial infall well before reaching the centre of the cluster.
\item
  For most massive galaxies, stripping radius falls below the gas disc
  scale length below $0.5$~$R_{vir}$ for the cluster.
  Thus we may expect to see some star-forming galaxies in the outer
  parts of a cluster, particularly if these are in their first passage
  through the cluster. 
\item
  In the case of non-radial infall, there is some gas left in the gas disc
  even after one crossing of the  galaxy: gas retention is stronger
  when the angular momentum of the infalling galaxy is larger for a 
  given initial speed. 
\item
  Ram pressure stripping is independent of redshift if we assume the
  worst case of the halo and galaxy forming at the same time.
  If the galaxy forms earlier, as is likely in most situations, more
  gas will be retained. For example, Eqn.\ref{eq:r_strip} will 
  be modified if the infalling
  galaxy collapsed at redshift $z_g$ and the halo into which it falls
  collapsed at $z_c$ with $z_g \geq z_c$:
\begin{equation}
\frac{R_{\rm strip}}{R_{\rm dg}} = \frac{R_{\rm ds}}{R_{\rm dg} +
  R_{\rm ds}} \left\{
\ln\left[\frac{2\pi
    G\Sigma_{0g}\Sigma_{0s}}{\rho(R_c)V^{2}(R_c)}\right] + 4 \ln\left[
  \frac{1 + z_g}{1 + z_c} \right] \right\}
\label{eq:r_stripz} 
\end{equation}
where the values of surface densities, density, velocity, etc. on the
right hand side of the equation are scaled to the present epoch.
Thus we see that $R_{\rm strip}$ can be large if the infalling galaxy
collapses at a higher redshift. 
\item
  We find that gas is stripped more easily for the larger spin parameter.
  In our model, the disc scale lengths are directly related to the
  spin parameter of their host halos: a larger spin parameter implies
  a larger disc scale length.
  This, in turn, implies that the available gas mass is distributed over
  a larger area.
  We have assumed that the ratio of gas mass (and stellar mass) is
  independent of the spin parameter.
  If the gas mass is in some manner correlated with the spin parameter
  of galaxy halos, then the variation seen in our model may not be correct. 
\item
  The variation of the initial velocity of the infalling galaxy does not
  matter much over the full orbit, though it does make a difference at
  early stages.
  This is due to the acceleration during infall dominating over any
  initial motion. 
\item
  Ram pressure stripping is sub-dominant in the group environment as
  compared to clusters.
  This is due to lower retention of hot gas in the intra-group medium
  and lower infall velocities due to shallower potential wells. 
\item
  This trend continues to lower scales as ram pressure stripping is
  less important in galaxy environments. 
\end{itemize}

There are several processes that may reduce the impact of ram pressure
stripping.
We have already discussed non-radial motions above.
One such process, which is expected to decrease the effect of ram 
pressure stripping is the formation of gas clouds or star-forming
regions in the galaxies.
Such regions of a significantly higher surface density of gas can
resist ram pressure much better than the idealised smooth exponential
disc considered here. This enables us to consider a 
non-axisymmetric distribution of gas and stars.
As an illustration of this we have considered a inhomogenous simulated 
Adaptive Mesh Refinement (AMR) code RAMSES and then subjecting it to
ram pressure striping algorithim. More than $50\%$ of gas is retained
in the disc due inhomogenities in the disc as shown in figures
\ref{fig:strip} and \ref{fig:mass_fraction}.
Thus star formation can continue in an infalling galaxy for about one
crossing time, if not longer.  At later times, feedback due to star
formation in remaining dense clouds will disperse the gas and star
formation rate will rapidly approach zero. 
We plan to explore on this extended model in detail in future. 

Ram pressure is expected to dominate in the regions where there is no
galaxy in the vicinity of the infalling galaxy.
Whenever there is another galaxy within the radius of influence of the
infalling galaxy, tidal stripping or galaxy harassment can also play an
important role.
Tidal stripping can remove all matter from the outer parts of the
infalling galaxy.
Considering time scales, we believe ram pressure stripping to be very
important in cluster halos while tidal stripping is likely to dominate
in galaxy halos.
In view of other processes, it is important to note here that our
analysis is valid only for first few orbits of the infalling galaxy.

The processes that contribute to significant deviations from the
worst case are:
\begin{itemize}
\item
  Non-radial orbits.
\item
  Early formation of infalling galaxies.
\item
  Inhomogeneous ISM in infalling galaxies.
\end{itemize}
As we have shown, each of these can contribute significantly and hence
in a realistic case the stripping radius can be an order of magnitude
higher than the estimates for worst case.

In case of ram pressure stripping in galaxy halos, there is one
scenario that can make the process very effective: if the
orbit of the infalling galaxy is such that it crosses the disc then
for a brief period, it encounters a high-density ambient medium.
This can be effective even if this crossing is in the outer parts of
the gas disc.

We have studied the process of ram pressure stripping, and we find that
it is a very efficient process for removing gas from discs of
infalling galaxies in clusters of galaxies.
We expect most galaxies to lose all their disc gas within a couple of
orbits around the centre of a cluster.
The process is less efficient in groups and clusters, and we expect to
see gas-rich galaxies in group and galaxy environments.

\section*{Acknowledgement}

We acknowledge the HPC@IISERM, used for computation during the project.  
We gratefully acknowledge the National Postdoctoral fellowship grant,
PDF/2015/000352, of Science and Engineering Research Board-Department
of Science and Technology, India for support during the fulfilment of
the work. The authors would like to thank Dr. Joakim Rosdahl, Centre de 
Recherche Astrophysique de Lyon, France for his guidance in the simulation 
of the isolated galaxy.
We would like to acknowledge Ms. Pooja Munjal for helping with fig.1.
This research has made use of NASA's Astrophysics Data
System Bibliographic Services. We would also like to thank the referee for his suggestions which improved the manuscript.

\bibliographystyle{mn2e}
\end{document}